\documentclass[
   a4paper,reqno]{article} 
\usepackage{amsmath, amsfonts,
   amssymb, amscd, amsthm, euscript}

\theoremstyle{definition}
\newtheorem*{define}{Definition}
\newtheorem*{lemma}{Lemma}
\theoremstyle{remark}
\newtheorem{rem}{Remark}
\newtheorem{example}{Example}

\DeclareFontFamily{OML}{cyr}{}
\DeclareFontShape{OML}{cyr}{m}{n}{
   <5> <6> <7> <8> <9> gen * wncyr
   <10> <10.95> <12> <14.4> <17.28> <20.74> <24.88> wncyr10
  }{}
\DeclareSymbolFont{rusletters}{OML}{cyr}{m}{n}
\DeclareSymbolFontAlphabet{\rusmath}{rusletters}
\DeclareMathSymbol\re{\rusmath}{rusletters}{"03}
\newcommand{\cEv}{\re}

\newcommand{\BBR}{\mathbb{R}}
\newcommand{\BBN}{\mathbb{N}}
\newcommand{\cC}{\mathcal{C}}
\newcommand{\cE}{\mathcal{E}}
\newcommand{\cEEL}{{\cE}_{\text{\textup{EL}}}}
\newcommand{\cH}{\mathcal{H}}
\newcommand{\cI}{\mathcal{I}}
\newcommand{\cL}{\mathcal{L}}
\newcommand{\cX}{{\EuScript X}}    
\newcommand{\cY}{{\EuScript Y}}    
\newcommand{\bE}{\mathbf{E}}
\newcommand{\gm}{\mathfrak{m}}
\newcommand{\gA}{\mathfrak{A}}

\newcommand{\veps}{\varepsilon}
\newcommand{\vph}{\varphi}
\newcommand{\dd}{\partial}
\newcommand{\Id}{{\mathrm d}}
\newcommand{\ID}{{\mathrm D}}
\newcommand{\IL}{{\mathrm L}}
\newcommand{\rmi}{{\mathrm i}}
\newcommand{\fnh}{{\text{\textup{FN}}}}

\DeclareMathOperator{\id}{id}
\DeclareMathOperator{\sym}{sym}
\DeclareMathOperator{\const}{const}
\DeclareMathOperator{\arcsinh}{arcsinh}
\DeclareMathOperator{\coker}{coker}

\newcommand{\tu}{\tilde{u}}
\newcommand{\tv}{\tilde{v}}
\newcommand{\tV}{\tilde{V}}
\newcommand{\lshad}{[\![}
\newcommand{\rshad}{]\!]}

\newcommand{\by}[1]{\textit{{#1}}}
\newcommand{\jour}[1]{\textit{{#1}}}
\newcommand{\vol}[1]{\textbf{{#1}}}
\newcommand{\book}[1]{\textrm{{#1}}}

\title
{Algebraic properties of Gardner's deformations for integrable systems}
\date{October 30, 2006; in final form December 18, 2006}

\author{A. V. Kiselev
\thanks{
Max Planck Institute for Mathematics,
Vivatsgasse 7, D-53111 Bonn, Germany; %
Institut des Hautes $\smash{\text{\'Etudes}}$ Scientifiques,
Le Bois\/--\/Marie 35, Route de Chartres,
F-\/91440 Bures\/-\/sur\/-\/Yvette, France.
\textit{E-mail}: \texttt{arthemy\symbol{"40}mpim-bonn.mpg.de}}%
\thanks{Proc.\ conf.\ `Nonlinear Physics: Theory and Experiment~IV'
(June 22~-- July~1, 2006) Gallipoli, Italy.}%
\thanks{Theor.\ Math.\ Phys.\ (2007) \vol{151}:3, accepted;
Preprint MPIM-162/2006 (Bonn).}%
}

\begin{document}

\maketitle


\begin{abstract}
An algebraic definition of Gardner's deformations for completely
integrable bi\/-\/Hamiltonian evolutionary systems is formulated. The
proposed approach extends the class of deformable equations and yields
new integrable evolutionary and hyperbolic Liouville\/-\/type systems.
An exactly solvable two\/-\/component extension of the Liouville
equation is found.
\end{abstract}

\textbf{Keywords:} {Gardner's deformations, integrable families,
adjoint systems, Hamiltonians, recurrence relations.}

\subsection*{Introduction}
We consider the problem of constructing Gardner's deformations of completely integrable bi\/-\/Hamiltonian evolutionary (super-)\/systems, see~\cite{Gardner,MathieuOpen,MathieuN=2,PamKale} or the review~\cite{KuperIrish} and references therein.
The essence of this procedure~\cite{Gardner} is that the generating vector\/-\/functions for the Hamiltonians of integrable systems solve auxiliary evolution equations which obey certain restrictions.
Then the deformations yield recurrence relations for densities of the Hamiltonians and result in parametric extensions of known systems.


In this paper we propose an algebraic definition of Gardner's deformations through diagrams~\eqref{GDiag}, which extends the class of deformable systems. Passing to the infinitesimal standpoint, we show that the Gardner deformations are inhomogeneous generalizations of the higher symmetries for~PDE. Then, applying recent geometric techniques~\cite{ThreeApp}, we interpret the Gardner deformation cohomology as a specification of the Cartan cohomology~\cite{Opava}. This invariant of integrable systems is related to the problem of classification of admissible Miura's transformations between them.

The method of~\cite{TMPhGallipoli} for reconstructing the substitutions using the ambient Euler\/-\/Lagrange Liouville\/-\/type systems~\cite{SokolovUMN} specifies several mechanisms that generate new integrable systems through the deformation problem.
First we recall that the relation between the integrals of Liouville\/-\/type systems and non\/-\/invertible differential substitutions results in new hyperbolic systems, see~\eqref{EqIra}  for an extension of the Liouville equation.
Secondly, the understanding of Liouville\/-\/type systems as diagrams~\eqref{LDef} assigns new integrable flows on the Cauchy data for their general solutions to symmetry hierarchies of these exactly solvable hyperbolic equations.
Further, we show that the classical Gardner deformation problem is simplified by extending the associated Liouville\/-\/type systems to parametric families.
Finally, we introduce the notion of adjoint completely integrable
systems such that the extensions 
interpolate between the original and the adjoint equations.
The exposition is illustrated by new deformations of the (Kaup\/-)\/Boussinesq systems and well\/-\/known extensions of scalar KdV\/-\/type equations.

The Gardner deformations are not a unique way to obtain extensions, modifications, and generalizations of KdV\/-\/type systems, see, e.g., \cite{DSViniti84,BorisovZykov} and references therein. On the other hand, the definition of the deformations proposed in this paper allows to extend the class of deformable systems (in particular, onto systems that are not represented in the form of a conservation law).
We also note that algebraic aspects of deformations for integrable systems were studied in~\cite{Dubrovin} in a related but different context.

\section{The classical Gardner deformation scheme}
\noindent%
Let us introduce some notation and recall necessary concepts. By definition, put~$f[u]=f(u,u_x,\ldots)$, here $u$ denotes the components of a vector ${}^t(u^1$, $\ldots$, $u^m)$, and let $\bE$  denote the variational derivative with respect to~$u$.
Consider the space of functionals $\cH=\int H[u]\,\Id x$ such that the
elements of the space are equivalent if they differ by the exact terms
with $H[u]=D_x(G[u])$. An operator~$A$ in total derivatives is
Hamiltonian if the bracket
$\{\cH_1,\cH_2\}=\langle\bE(\cH_1),A(\bE(\cH_2))\rangle$ endowes the
space of functionals with a Lie algebra structure. Let the Hamiltonian
operators $A_1$, $A_2$ be compatible, that is, their arbitrary linear
combination be Hamiltonian again. Suppose further that there are $m$
infinite sequences $\{\cH^i_j$; $i=1,\ldots,m$; $j\in\BBN\}$ of the
functionals that satisfy the relation
\begin{equation}\label{MagriRel}
A_2(\bE(\cH^i_j))=A_1(\bE(\cH^i_{j+1}))
\end{equation}
for all~$i,j$. Fix some $i$ and $j$ and consider the evolutionary system $\cE=\{u_t=A_1(\bE(\cH^i_{j+1}))=A_2(\bE(\cH^i_j))\}$. Then $\cE$ is a completely integrable system; it admits $m$ infinite sequences of commuting symmetries $u_{t_k^l}=A_1(\bE(\cH^l_k))$
assigned to the Hamiltonians. All the densities $H^l_k$ are conserved on $\cE$: $D_t(H^l_k)\doteq D_x(\ldots)$.

\begin{rem}
For the sake of transparency, we do not use the purely algebraic definition of a completely integrable bi-Hamiltonian system specified by a pair of compatible Hamiltonian operators such that the Poisson cohomology~\cite{Getzler,Lstar} with respect to one of them is trivial, implying that relation~\eqref{MagriRel} can always be resolved for~$\cH^i_{j+1}$. Therefore we postulate the existence of the Hamiltonians.
\end{rem}

Let~$\cE=\{F\equiv u_t-f[u]=0\}$ be a completely integrable system.
Suppose~$\cE_\veps=\{F_\veps\equiv\tu_t-f_\veps([\tu],\veps)\}$ is a
deformation of~$\cE$ such that for any point~$\veps\in\cI$ 
of an interval~$\cI\subset\BBR$ there is the contraction
$\gm_\veps=\{u=u([\tu],\veps)\}\colon\cE_\veps\to\cE$.
Further, assume that~$f_\veps\in\text{im}\,D_x$ for all~$\veps$.
Then the pair~$\smash{\bigl(\cE_\veps,\gm_\veps\bigr)}$ is the
\emph{Gardner deformation} for~$\cE$ in the classical sense of~\cite{Gardner,MathieuOpen,KuperIrish}.\label{pClass}
Expanding the generating vector\/-\/function~$\tu=\sum_{n=0}^{+\infty}\tu_n\cdot\veps^n$ and~$\gm_\veps$ in~$\veps$, we
obtain the recurrence relations for infinite sequences of
the term\/-\/wise conserved densitied~$\tu_n[u]$
for~$\cE$, the sequences containing the Hamiltonians of higher flows for~$\cE$.

\begin{example}[\textup{\textmd{\cite{Gardner,KuperIrish}}}]\label{ExKdVe}
The extended KdV equation
\begin{subequations}\label{KdVDef}
\begin{align}\label{KdVe}
\tu_t&=-\tfrac{1}{2}\tu_{xxx}+3\tu\tu_x+3\veps^2\tu^2\tu_x\\
\intertext{is reduced to the KdV equation $u_t=-\tfrac{1}{2}u_{xxx}+3uu_x$ by the Miura contraction}
\gm_\veps&=\smash{\bigl\{u=\tu\pm\veps\tu_x+\veps^2\tu^2\bigr\}}.
\label{KdVeKdV}
\end{align}
\end{subequations}
Note that the KdV equation is invariant w.r.t.\ the Galilean transformation $x\mapsto x+3\lambda t$, $t\mapsto t$, $u\mapsto u+\lambda$.
Consider the modified KdV equation
\begin{equation}\label{mKdV}
v_t=-\tfrac{1}{2}v_{xxx}+3v^2v_x
\end{equation}
and Miura's substitution $u=v^2\pm v_x$ into~KdV. Put $v=\veps\tu+(2\veps)^{-1}$; this yields the contraction $u=\left(\tu\pm\veps\tu_x+\veps^2\tu^2\right)+\frac{1}{4\veps^2}=\gm_\veps+\lambda$ from~\eqref{KdVe} to~KdV, here we put $\lambda=\frac{1}{4\veps^2}$ and absorb the shift of $u$ by the motion along~$x$.
We see that the Galilean invariance of KdV generates the family~\eqref{KdVe} by using the modified KdV equation~\eqref{mKdV}.

Next, equation~\eqref{mKdV} is extended by
\begin{equation}\label{mKdVe}
\tv_t=-\tfrac{1}{2}\tv_{xxx}+3\tv^2\tv_x
 +D_x\left({3\veps^2\tv\tv_x^2}\bigr/{(1+4\veps^2\tv^2)}\right)
\end{equation}
such that the contraction $\gm_\veps$ is $v=\tv+\veps\tv_x/\sqrt{1+4\veps^2\tv^2}$.
The extended KdV~\eqref{KdVe} and the extended
modified KdV equation~\eqref{mKdVe} are related by the substitution
\[
\tu=\tfrac{1}{2\veps^2}\bigl(\sqrt{1+4\veps^2\tv^2}-1\bigr)
 +{\tv_x}\bigr/{\sqrt{1+4\veps^2\tv^2}}.
\]
\end{example}

\begin{example}\label{ExKBe}
The extension~$\cE_\veps$ of the Kaup\/--\/Boussinesq equation
\begin{equation}\label{KB}
\cE=\{u_t=uu_x+v_x,\quad v_t=(uv)_x+ u_{xxx}\}.
\end{equation}
is the system
\begin{subequations}\label{GDKB}
\begin{align}
\tu_t&=\tu\tu_x+\tv_x
   + \veps\cdot\bigl(\tu\tu_{xx}+\tu_x^2+{(\tu\tv)}_x\bigr),\notag\\
\tv_t&={(\tu\tv)}_x + \tu_{xxx}
   - \veps\cdot\bigl(2\tu_x\tu_{xx}+\tu\tu_{xxx}+\tu_x\tv_x+\tu\tv_{xx}
      - \tv\tv_x\bigr).\label{KBe}\\
\intertext{The contraction $\cE_{\veps}\to\cE$ is given through}
u&=\tu+\veps\cdot\bigl(\tu_x+\tv\bigr),\quad
v=\tv+\veps\cdot\bigl(\tu\tu_x+\tu_{xx}+\tu\tv+\tv_x\bigr).
\label{MiuraKBeKB}
\end{align}
\end{subequations}
The recurrence relations upon densities of the Hamiltonian functionals
for~\eqref{KB} are
\begin{align}
\tu_0&=u,\qquad
\tv_0=v,\qquad
\tu_k=-D_x(\tu_{k-1})-\tv_{k-1},\notag \\
\tv_k&=-D_x^2(\tu_{k-1}) - D_x(\tv_{k-1})
   - \sum_{\ell+m=k-1}
      \bigl[\tu_\ell D_x(\tu_m) + \tu_\ell \tv_m\bigr],\quad k>0.
      \label{KBHamRec}
\end{align}
Using the Gardner deformation~\eqref{GDKB} of the Kaup\/--\/Boussinesq equation, we shall construct a new exactly solvable Liouville\/-\/type system~\eqref{EqIra} and we shall also obtain the bi\/-\/Hamiltonian adjoint system~\eqref{AdjBous}.
\end{example}

\section{Definition of Gardner's deformations through diagrams}%
\label{SecDefGDef}
\noindent%
Summarizing the properties of Gardner's deformations for (non-)\/modified integrable systems~\cite{MathieuOpen,KuperIrish}, we observe that the classes of differential functions which determine the deformations are enlarged if there is a Miura\/-\/type transformation~$\tau$ that acts from the system at hand (e.g., irrational dependencies appear instead of polynomials, see Example~\ref{ExKdVe}).
Clearly, the classical deformations~$(\cE_\veps,\gm_\veps)$ do not exist for equations which are not in divergent form.
However, suppose two completely integrable systems $\cE$ and~$\cE'$ and their higher symmetries are connected by a Miura\/-\/type
transformation $\tau\colon\cE'\to\cE$; then their Hamiltonians are correlated by~$\tau^*\colon\cH[u]\mapsto\cH[u[v]]$, where~$u=\tau[v]$. Therefore the problem of reconstructing a recurrence relation for the Hamiltonians of the hierarchy for~$\cE'$ is naturally reduced to a Gardner's deformation of~$\cE$, provided that $\cE'$~admits only a finite number of its own Hamiltonians that do not belong to the image of~$\tau^*$. In particular, the problem of Gardner's deformation for the extension~$\cE_\veps$ is tautological (nevertheless, see~\cite{KuperIrish} for two examples). Now we propose a definition of the Gardner deformations that incorporates the above reasoning.

\begin{define}
The \emph{Gardner deformation} of a completely integrable evolutionary system~$\cE'$ is the diagram
\begin{equation}\label{GDiag}
\begin{CD}
\cE' @>{\tau}>> \cE @<{\gm_\veps}<< \cE_\veps,
\end{CD}
\end{equation}
where $\tau$~is a Miura\/-\/type transformation and
the contraction $\gm_\veps=\{u=u([\tu],\veps)\}$
maps the extension $\cE_\veps=\{F_\veps\equiv\tu_t-f_\veps([\tu],\veps)$,
$f_\veps\in\text{im}\,D_x\}$ to~$\cE$.
\end{define}

\begin{rem}
The standard approach~\cite{Gardner,MathieuN=2,KuperIrish,MathieuN1} corresponds to the identity transformation~$\tau=\id$.
Next, the transformation~$\tau$ may be invertible meaning that the coordinate system which determines~$\cE'\simeq\cE$ is `inconvenient.'
Indeed, an invertible coordinate transformation can destroy the form
$u=\tu+\veps\cdot[\ldots]$ of the contraction
$\gm_\veps\colon\cE_\veps\to\cE$, and the curve~$\cE_\veps$ will no longer contain~$\cE$ at~$\veps=0$.
It would be of interest to formalize the obstruction defined on the set of all coordinate systems on the equation~$\cE'$ such that the vanishing of the obstruction implies the form $u=\tu+\veps\cdot[\ldots]$ of~$\gm_\veps$.
\end{rem}

\begin{example}\label{ExBroer}
The Kaup\/--\/Boussinesq equation~\eqref{KB} is mapped
by the substitution $w=v-u_x$ to the Kaup\/--\/Broer system
\begin{equation}\label{KBroer}
u_t=u_{xx}+uu_x+w_x,\qquad w_t=(uw)_x-w_{xx}.
\end{equation}
Hence, using equations~\eqref{GDKB},
we obtain the Gardner deformation for~\eqref{KBroer};
the contraction from~\eqref{KBe} to~\eqref{KBroer} is
\[
u=\tu+\veps\cdot(\tu_x+\tv),\qquad
w=\bigl[\tv-\tu_x\bigr]+\veps\tu\cdot(\tu_x+\tv).
\]
\end{example}

Another example of a deformation for the $N=1$ supersymmetric KdV
such that the equation~$\cE_\veps$ is brought to a conservation law form by an invertible substitution is considered in~\cite{Kiev2005}.

\begin{rem}
Diagram~\eqref{GDiag} means that Gardner's deformations are in a sense dual to the B\"acklund transformations
$\smash{\cE'\xleftarrow{\tau}\cE \xrightarrow{\tau'}\cE''}$
between differential equations $\cE'$ and~$\cE''$, see~\cite{Opava}.
Clearly, discrete symmetries~$\sigma$ of the extensions~$\cE_\veps$ induce the B\"acklund autotransformations $\cE\xleftarrow{\gm_\veps}\cE_\veps\xrightarrow{\gm_\veps\circ\sigma}\cE$ for the equation~$\cE$. For example, the alteration of signs in contraction~\eqref{KdVeKdV} combined with elimination of the variable~$\tu$ provides the well\/-\/known one\/-\/parametric B\"acklund autotransformation for the Korteweg\/--\/de Vries equation, see~\cite{KuperIrish}.
\end{rem}


Now we claim that the Gardner deformations are
inhomogeneous generalizations of the higher symmetries. This is readily seen by passing to the infinitesimal standpoint.

\begin{lemma}\label{LemGSym}
Let~$\veps_0\in\cI$ such that there is the
mapping~$\gm_{\veps_0}\colon\cE_{\veps_0}\to\cE=\{F=0\}$;
put $\vph(\veps)=\dd\gm_\veps/\dd\veps$.
Then the contraction~$\gm_\veps$
satisfies the relation which holds by virtue ($\doteq$)
of~$\cE_{\veps_0}=\{F_{\veps_0}=0\}$,
\begin{equation}\label{Inhomog}
\cEv_{\vph(\veps_0)}(F) \doteq
{\cEv_{\dd F_\veps/\dd\veps}(\gm_\veps)
\Bigr|}_{\veps=\veps_0}.
\end{equation}
Equation~\eqref{Inhomog} also holds for deformations of
non\/-\/evolutionary systems.
\end{lemma}

The infinitesimal approach leads to the Gardner deformation
cohomology, which are determined using the standard
techniques~\cite{ThreeApp,Opava,Gerstenhaber}. Let us first recall the
notion of Cartan's cohomology (see~\cite{Opava} for details). Let
$\cE$~be a differential equation and $U(\cE)$ be the Cartan connection
form on it; in coordinates, we have
$U(\cE)=\sum_\sigma\Id_\cC(u_\sigma)\otimes\dd/\dd u_\sigma$, where
$u_\sigma$ runs through all derivatives of~$u$ that parameterize~$\cE$
and $\Id_\cC\colon u_\sigma\mapsto \Id u_\sigma-u_{\sigma+1_i}\,\Id x^i$
is the Cartan differential. Let $\mu=\deg\Omega$ if
$\Omega\in\ID(\Lambda^\mu(\cE))$ is a form\/-\/valued derivation, let
$\rmi_\Omega\colon\Lambda^k(\cE)\to\Lambda^{k+\deg\Omega-1}(\cE)$ be
the inner product,
$\IL_\Omega=[\rmi_\Omega,\Id]\colon\Lambda^k(\cE)\to
\Lambda^{k+\deg\Omega}(\cE)$ be the Lie derivative,
and $[\![\cdot,\cdot]\!]^{\fnh}$ be the Fr\"olicher\/--\/Nijenhuis
bracket such that $[\![\Omega,\Theta]\!]^\fnh(f)=
\IL_\Omega(\Theta(f))-(-1)^{\mu\nu}\IL_\Theta(\Omega(f))$ for any
$\Omega\in\ID(\Lambda^\mu(\cE))$, $\Theta\in\ID(\Lambda^\nu(\cE))$,
and $f\in C^\infty(\cE)$. A cumbersome coordinate formula for
$[\![\cdot,\cdot]\!]^\fnh$ is given
in~\cite{Opava,DeformLiou}.
The Cartan connection $U(\cE)$ satisfies the flatness condition
$[\![U(\cE),U(\cE)]\!]^\fnh=0$ implying that
$\dd_\cC=[\![\cdot,U(\cE)]\!]^\fnh$ is a differential. The Cartan
cohomology w.r.t.\ $\dd_\cC$ contains symmetries of~$\cE$ in the zeroth
term, recursions for~$\cE$ in the first term,
etc.~\cite{ThreeApp,Opava}.

This algebraic construction can be reformulated~\cite{ThreeApp}, in particular, for the families of equations~$\cE_\veps$ that admit the contractions
$\gm_\veps\colon\cE_\veps\to\cE$; then the Cartan cohomology describes (shadows of, see~\cite{Opava}) symmetries that are nonlocal w.r.t.\ the underlying equation~$\cE$ and do depend on the transformations~$\gm_\veps$.

Now consider all possible deformations $U_\veps(\cE_\veps)$ of the connection form; here we assume that for any~$\veps$ the Cartan distribution at every point of~$\cE_\veps$ remains isomorphic to the one at the corresponding point of~$\cE$ under some projection,
which is not necessarily an explicit globally
defined substitution~$\gm_\veps$.
The exact deformations are determined by (shadows of) symmetries $X_{\veps_0}$ for~$\cE$ that can not be lifted onto the corresponding~$\cE_{\veps_0}$ and hence do not leave it invariant but propagate it to a family along~$\veps$; then the Cartan connection $U_\veps$ on $\cE_\veps$ evolves by the equation~\cite{ThreeApp,Opava}
\begin{equation}\label{EvolveSE}
\frac{\Id U_\veps}{\Id\veps}=[\![X_\veps,U_\veps]\!]^\fnh,
\end{equation}
see~\cite{DeformLiou} for an illustration. However, \emph{all} deformations of $U_\veps$ are closed w.r.t.\ $\dd_\cC$, that is, we always have $[\![\frac{\Id U_\veps}{\Id\veps},U_\veps]\!]^\fnh=0$.
Therefore this Cartan's cohomology is infinite and hardly tractable if we omit the requirement that the contractions~$\gm_\veps\colon\cE_\veps\to\cE$ exist at any~$\veps$.

Now we formulate the notion of Gardner's deformation cohomology for a completely integrable system~$\cE$.
Consider all families $\{\cE_{\veps_\alpha}\}$ of systems that admit the Miura\/-\/type transformations $\gm_{\veps_\alpha}\colon\cE_{\veps_\alpha}\to\cE$. For each $\alpha$ the family $\gm_{\veps_\alpha}$ determines the Cartan deformation cohomology according to the standard scheme~\cite{ThreeApp,Opava} such that in each case all constructions depend on the correspondence~$\gm_{\veps_\alpha}$. However, the connection forms~$U_{\veps_\alpha}$ and the differentials $\smash{\dd_\cC^{(\alpha)}}$ are defined according to a unique scheme, which can be further extended onto the
sum $\bigoplus_\alpha \gm_{\veps_\alpha}$ of the coverings~$\cE_{\veps_\alpha}\to\cE$. Thus, using the Cartan connection~$\tilde{U}$ on the sum, 
we obtain the differential $\dd_\cC=[\![\cdot,\tilde{U}]\!]^\fnh$.
Consequently, the deformation cohomology is then defined w.r.t.\ $\dd_\cC$ in agreement with~\cite{ThreeApp}; this is the Gardner cohomology of the completely integrable system~$\cE$.

Their essential distinction from the Cartan cohomology is that the Gardner deformations, even if understood in the classical sense~$(\cE_\veps,\gm_\veps)$, impose a severe restriction upon the admissible extensions~$U_\veps$ through the Lemma;
namely, the existence of the contractions~$\gm_\veps$ must be preserved. The classes of exact
deformations\footnote{We differ the exact deformations, which are described by the mechanism~\eqref{EvolveSE}, and the trivial deformations, which generate only a finite number of nontrivial conserved densities via the
recurrence relations.
A trivial deformation of the `minus' Kaup\/--\/Boussinesq equation was found in~\cite{PamKale}. Hence \eqref{KdVDef}~is an example of a non\/-\/trivial exact deformation, and \eqref{GDKB} is non\/-\/trivial and not exact.}
are counted by the symmetries of~$\cE$ that do not lift onto~$\cE_\veps$ but propagate it to a family such that relation~\eqref{Inhomog} holds. As for the cocycles, we conjecture that the Gardner cohomology for KdV\/-\/type systems originating from the Lie algebras~\cite{DSViniti84} are finite\/-\/dimensional, that is, these systems admit a finite number of recurrence relations between the Hamiltonians. The conjecture means that the Gardner cohomology is an invariant of integrable systems which can be helpful in the classification problems.

Example~\ref{ExKdVe} describes an exact deformation~\eqref{KdVDef} of the KdV equation by using its Galilean symmetry. This symmetry yields the deformations of many KdV\/-\/type systems~\cite{KuperIrish} but not of all of them~\cite{MathieuN1}. Indeed, this symmetry is already lost by the modified KdV equation which is extended by~\eqref{mKdVe}.
We also recall that under $\tau\colon\cE'\to\cE$ the algebra $\sym\cE'$ is embedded in~$\sym\cE$; hence if the inclusion $\tau_*\colon\sym\cE'\subseteq\sym\cE$ is strict and the symmetry of~$\cE$ that generates the family~$\cE_\veps$ belongs to~$\coker\tau_*$, then the arising Gardner deformation for~$\cE$ can not be transferred to~$\cE'$ within the classical scheme.
Generally, the presence of Gardner's cohomology disclaims an \emph{ad hoc} principle that a symmetry of~$\cE$ can be always used for generating the family~$\cE_\veps$.
Counter\/-\/examples of the cocycles which are not coboundaries are given, e.g.,
by the modified KdV equation (see Example~\ref{ExKdVe}), by the $N=1$ super\/-\/KdV equation~\cite{MathieuN1}, the Kaup\/--\/Boussinesq equation in Example~\ref{ExKBe}, or by the Boussinesq system.

\begin{example}\label{ExBouse}
There are two Gardner's deformations of the Boussinesq equation
$u_t=v_x$, $v_t=u_{xxx}+uu_x$.
The extended Boussinesq equations~$\cE_\veps^{\pm}$ are
\begin{subequations}\label{BousDef}
\begin{align}
\tu_t&=\tv_x+\varepsilon^3\cdot\bigl(
   \tu_x\tu_{xx}\pm\tu_{xx}\tv\pm\tu_x\tv_x+\tv\tv_x\bigr),\notag\\
\tv_t&=\tu_{xxx}+\tu\tu_x - \varepsilon^3\cdot\bigl(
   \tu_{xxx}\tv + 2\tu_{xx}\tv_x + \tu_x\tv_{xx}
   \pm \tu_{xx}^2 \pm \tu_x\tu_{xxx} \pm \tv_x^2 \pm \tv\tv_{xx}\bigr).\label{Bouse}\\
\intertext{The respective
contractions~$\gm^{\pm}_\veps\colon\cE^{\pm}_\veps\to\cE$
are given through}
u&=\tu\mp2\varepsilon \tu_x + 2\veps^2\cdot(\tu_{xx}\pm\tv_x)
   + \veps^3\cdot(\tu\tv\pm\tu\tu_x),\notag\\
v&=\tv\mp2\veps \tv_x + 2\veps^2\cdot(\tv_{xx}\pm\tu_{xxx}\pm\tu\tu_x)
   + \veps^3\cdot\Bigl(\tfrac{1}{3}\tu^3 + \tv^2 + \tu\tu_{xx}
     \pm \tv\tv_x \pm \tu_x\tv \Bigr)\notag\\
{}&{}\qquad
   \mp 2\veps^4\cdot(\tu_x\tu_{xx}\pm\tu_x\tv_x\pm\tu_{xx}\tv+\tv\tv_x)
   + \veps^6\cdot\Bigl(\tfrac{1}{3}\tv^3 \pm \tfrac{1}{3}\tu_x^3
     + \tu_x^2\tv \pm \tu_x\tv^2\Bigr).\label{BouseBous}
\end{align}
\end{subequations}
The list of classical symmetries of the Boussinesq equation is exhausted by the translations and the dilation. None of them generates the family~\eqref{Bouse} by propagating the systems $\cE_{\veps_0}$ at any~$\veps_0$, and hence the deformation~\eqref{BousDef} is not exact.

Expanding the fields $\tu$, $\tv$ and the
contractions~$\gm^{\pm}_\veps$ in~$\veps$,
we obtain the recurrence relations for the two sequences
of Hamiltonians for the Boussinesq equation
(thus we generalize a result of~\cite{KuperIrish}, where one relation for only one sequence was obtained):
\begin{align*}
\tu_0&=u, \quad \tv_0=v, \quad \tu_1=\pm 2u_x,\\
\tv_1&=\pm 2v_x, \quad \tu_2= 2u_{xx} \mp 2v_x, \quad
\tv_2= 2v_{xx} \mp 2u_{xxx} \mp 2uu_x,\\
\tu_n&=\pm 2D_x(\tu_{n-1}) - 2D_x^2(\tu_{n-2}) \mp 2D_x(\tv_{n-2})
   +\sum_{k+\ell=n-3}\Bigl[-\tu_k\tv_\ell \mp \tu_kD_x(\tu_\ell)\Bigr],\\
   {}&\qquad{}\qquad n\geq3,\\
\tv_3&=\pm 2D_x(\tv_2) \mp 2D_x^3(\tu_1) - 2D_x^2(\tv_1)
   \mp \left[uD_x(\tu_1) + \tu_1u_x\right] \\
  {}&\qquad{}
   - \frac{1}{3}u^3 - v^2 - uu_{xx} \mp uv_x \mp u_xv,\\
\tv_n&=\pm 2D_x(\tv_{n-1}) \mp 2D_x^3(\tu_{n-2}) - 2D_x^2(\tv_{n-2}) \\
  {}&\qquad{}
   \mp \sum_{k+\ell=n-2}2\tu_kD_x(\tu_\ell)
   - \sum_{k+\ell+m=n-3}\frac{1}{3}\tu_k\tu_\ell \tu_m\\
  {}&\qquad{}
   + \sum_{k+\ell=n-3}\Bigl[- \tv_k\tv_\ell - \tu_kD_x^2(\tu_\ell)
      \mp \tu_kD_x(\tv_\ell) \mp D_x(\tu_k)\tv_\ell\Bigr] \\
  {}&\qquad{}
   + \sum_{k+\ell=n-4}2\cdot\Bigl[\pm D_x(\tu_k)D_x^2(\tu_\ell)
      + D_x(\tu_k)D_x(\tv_\ell) + D_x^2(\tu_k)\tv_\ell
      \pm \tv_kD_x(\tv_\ell)\Bigr],\\
  {}&\qquad{}\qquad n=4,5,\\
\tv_n&=\pm 2D_x(\tv_{n-1}) \mp 2D_x^3(\tu_{n-2}) - 2D_x^2(\tv_{n-2}) \\
  {}&\qquad{}
   \mp \sum_{k+\ell=n-2}2\tu_kD_x(\tu_\ell)
   - \sum_{k+\ell+m=n-3}\frac{1}{3}\tu_k\tu_\ell \tu_m\\
  {}&\qquad{}
   + \sum_{k+\ell=n-3}\Bigl[- \tv_k\tv_\ell - \tu_kD_x^2(\tu_\ell)
      \mp \tu_kD_x(\tv_\ell) \mp D_x(\tu_k)\tv_\ell\Bigr] \\
  {}&\qquad{}
   + \sum_{k+\ell=n-4}2\cdot\Bigl[\pm D_x(\tu_k)D_x^2(\tu_\ell)
      + D_x(\tu_k)D_x(\tv_\ell) + D_x^2(\tu_k)\tv_\ell
      \pm \tv_kD_x(\tv_\ell)\Bigr] \\
  {}&\qquad{}
   + \sum_{k+\ell+m=n-6}\Bigl[-\frac{1}{3}\tv_k\tv_l\tv_m
      \mp \frac{1}{3}D_x(\tu_k)D_x(\tu_\ell)D_x(\tu_m) \\
  {}&\qquad{}\qquad
      - D_x(\tu_k)D_x(\tu_\ell)\tv_m + D_x(\tu_k)\tv_\ell \tv_m\Bigr],
  \qquad n\geq6.
\end{align*}
The ambiguity of signs of the differential terms does not affect the
nontrivial conserved densities $\tu_{3k}$ and $\tv_{3k}$.
The densities with subscripts $3k+1$, $3k+2$
are trivial for all~$k\geq0$.
\end{example}

We conclude that the vanishing of Gardner's cohomology is an obstruction to existence of a recurrence relation between the Hamiltonians of the hierarchy if, further, there are no Gardner's deformations obtained using symmetries of the initial system (in particular, if it has no classical symmetries at all except the translations). The task to calculate the Gardner cohomology can be addressed by various techniques (in a similar problem, see an estimation of the Poisson cohomology through the de Rham cohomology in~\cite{Getzler}). A prerequisite to this problem is the classification of Miura's substitutions~$\gm_{\veps_\alpha}$ into the system at hand with respect to~$\alpha$. The general geometric problem of constructing the Miura transformations between integrable systems is addressed in the next section.

\section{Gardner's deformations and the ambient Liouville\/-\/type systems}\label{SecLiou}
\noindent%
In this section we demonstrate how the problem of constructing
Gardner's deformations~\eqref{GDiag} can be simplified by
the use of hyperbolic Liouville\/-\/type systems,
which also admit interpretation~\eqref{LDef} through diagrams.
First we recall the method~\cite{TMPhGallipoli} of representing the
hierarchies as commutative Lie subalgebras of symmetries of
Euler\/--\/Lagrange
Liouville\/-\/type systems~$\cEEL$ that specify the substitutions
$\tau\colon\cE'\to\cE$ within diagram~\eqref{GDiag}.
The evolutionary systems~$\cE$ in the image of these transformations
are always symmetries of the multi\/-\/component wave equations, and
Examples~\ref{ExKdVe}--\ref{ExBouse} show that their Gardner
deformations then obey the classical scheme.
We also demonstrate that the ambient systems~$\cEEL$ can be propagated to families~$\cE_\IL^\veps$ such that the extensions remain Liouville\/-\/type; the contractions $\gm_\veps\colon\cE_\veps\to\cE$ are then induced by the transformations $\cE_\IL^\veps\to\cEEL$.
Thus the  Liouville\/-\/type systems~$\cEEL$ resolve both arrows in diagram~\eqref{GDiag} and, reciprocally, the deformation problem leads to new integrable systems.
Other integrable evolutionary equations are obtained by lifting the
hierarchies onto the general solutions of~$\cEEL$. In what follows, we
construct an exactly solvable two\/-\/component generalization of the
Liouville equation and we find a Liouville\/-\/type parametric
extension of this equation; also, we recall that the lift of the
modified KdV symmetry of the Liouville equation onto its solution
is the Schwarz\/-\/KdV equation.

Now we consider an example whose nature is essentially general, applying the method of~\cite{TMPhGallipoli} for
constructing Miura's substitution~$\tau$
for the evolutionary system~\cite{ConstCurv}
\begin{equation}\label{m2KB}
a_t=\tfrac{1}{2}\bigl(a^2b-2a_xb-4b_{xx}\bigr)_x,\qquad
b_t=\tfrac{1}{2}\bigl(a(b^2-1)+2bb_x\bigr)_x.
\end{equation}
Namely, we obtain the Gardner deformation for~\eqref{m2KB} not including it in a one\/-\/parametric family, that is, trying to perform the classical scheme, but using the algebraic definition of the deformations as diagrams~\eqref{GDiag}. Hence we first construct substitution
~\eqref{m2KBKB} that transforms~\eqref{m2KB} to system~\eqref{KB}
whose deformation is already known. To this end, we potentiate the
hierarchy of~\eqref{m2KB} using~\eqref{pm2KBm2KB} and then we represent
higher symmetries of system~\eqref{pm2KB} as a commutative Lie
subalgebra of Noether's symmetries of the Euler\/--\/Lagrange
Liouville\/-\/type system~\eqref{EqIra}
whose integrals~\eqref{Integrals} determine the required
substitution~\eqref{m2KBKB}. The two\/-\/component hyperbolic
system~\eqref{EqIra} is exactly solvable, and we construct its general
solution that depends on four arbitrary functions.

System~\eqref{m2KB} is bi\/-\/Hamiltonian
; its first structure is $
\left(\begin{smallmatrix}0 & D_x\\ D_x & 0\end{smallmatrix}\right)$.
Hence we introduce the potentials $A$, $B$ such that the adjoint linearization $\bigl(\ell_{a,b}^{(A,B)}\bigr)^*$ of $a,b$ with respect to $A,B$ is proportional to the first structure.
So we set
\begin{equation}\label{pm2KBm2KB}
a=\tfrac{1}{2}B_x,\qquad b=\tfrac{1}{2}A_x.
\end{equation}
The new variables satisfy the Hamiltonian system
\begin{equation}\label{pm2KB}
A_t=\tfrac{1}{2}A_xA_{xx}+\tfrac{1}{2}\left(\tfrac{1}{4}A_x^2-1\right)B_x,\qquad
B_t=-2A_{xxx}+\tfrac{1}{8}A_xB_x^2-\tfrac{1}{2}A_xB_{xx}.
\end{equation}
We remark that systems~\eqref{m2KB} and~\eqref{pm2KB} can be now cast to the canonical form,
\[
A_t=\delta\cH/\delta a,\quad a_t=-\delta\cH/\delta A;\qquad
B_t=\delta\cH/\delta b,\quad b_t=-\delta\cH/\delta B,
\]
where the Hamiltonian functional $\cH=\int H[a,b]\,\Id x$ has the
density
$H=\tfrac{1}{32}A_x^2B_x^2+\tfrac{1}{4}A_xA_{xx}B_x+\tfrac{1}{2}A_{xx}^2-\tfrac{1}{8}B_x^2$.

The principal idea of the algorithm of~\cite{TMPhGallipoli} is the construction of the Lagrangian $\cL=-\iint[a\cdot A_y+b\cdot B_y+H(A,B)]\Id x\,\Id y$ such that the potential system~\eqref{pm2KB} and the hierarchy of its higher flows are Noether's symmetries of the hyperbolic Euler\/--\/Lagrange equations~$\cEEL=\{\bE(\cL)=0\}$. A tedious but straightforward calculation leads to the following extension of the Liouville equation,
\begin{equation}\label{EqIra}
A_{xy}=-\tfrac{1}{8}A\exp\bigl(-\tfrac{1}{4}B\bigr),\qquad
B_{xy}=\tfrac{1}{2}\exp\bigl(-\tfrac{1}{4}B\bigr).
\end{equation}
System~\eqref{EqIra} is Liouville\/-\/type: its integrals are
\begin{equation}\label{Integrals}
I_1=-\tfrac{1}{4}a^2-a_x,\qquad I_2=ab+2b_x
\end{equation}
such that $\bar{D}_y(I_{i})\doteq0$ on~\eqref{EqIra}, $i=1,2$.
The Liouville\/-\/type system~\eqref{EqIra} is not contained in the classification
lists~\cite{Demskoi} since it is triangular.

Further, consider the integrals
\begin{equation}\label{m2KBKB}
u=I_2, 
\qquad
v=I_1+\tfrac{1}{4}I_2^2       
\end{equation}
such that the second of them is not minimal. Calculating their dynamics along evolution equation~\eqref{m2KB}, we obtain the Kaup\/--\/Boussinesq system~\eqref{KB}. Thus we recover the
substitution~\eqref{m2KBKB} to~\eqref{KB} from the twice\/-\/modified equation~\eqref{m2KB}, see~\cite{ConstCurv}. The Gardner deformation for the Kaup\/--\/Boussinesq system~\eqref{KB} is already
described in Example~\ref{ExKBe}, hence the recurrence relations for the Hamiltonians of higher symmetries of systems~\eqref{m2KB} and~\eqref{pm2KB} are inherited from~\eqref{KBHamRec} by using substitutions~\eqref{pm2KBm2KB} and~\eqref{m2KBKB}.

Clearly, the linear hyperbolic extension~\eqref{EqIra} of the scalar Liouville equation is integrable. However, the ambient system~\eqref{EqIra} admits two independent integrals~\eqref{Integrals} and therefore its general solution can be obtained explicitly. This is done as follows (other techniques for solving equations of this class are exposed in~\cite{SokolovUMN}). First recall that the well\/-\/known solution of the Liouville equation in system~\eqref{EqIra} is
\[
B(x,y)=-4\ln\frac{-16\cX'(x)\cY'(y)}{Q^2(\cX(x)+\cY(y))},
\]
where $\cX(x)$ and $\cY(y)$ are arbitrary functions and $Q\in\{\id,\sin,\sinh\}$. Hence the other equation in~\eqref{EqIra} acquires the form
\begin{equation}\label{Hills}
A_{xy}=2A\cdot\frac{\cX'(x)\cY'(y)}{Q^2(\cX(x)+\cY(y))}.
\end{equation}
Performing the transformation $\tilde{x}=\cX(x)$ and $\tilde{y}=\cY(y)$ and from now on omitting the tilde signs, we arrive at the linear equation
\begin{equation}\tag{\ref{Hills}${}'$}\label{HillsTr}
A_{xy}={2A}\bigr/{Q^2(x+y)}.
\end{equation}
Secondly, let $f(x)$ and $g(y)$ be arbitrary functions.
The minimal integral~$I_2$ in~\eqref{Integrals} belongs to the kernel of the derivative~$\bar{D}_y$ restricted onto~\eqref{EqIra}, therefore we set $I_2\mathop{{:}{=}}f(x)$ and consider the arising linear inhomogeneous ordinary differential equation with respect to~$b=\tfrac{1}{2}A_x$.
Solving it and symmetryzing the intermediate result in $x$ and~$y$, we finally obtain the three solutions $A(x,y)$ that correspond to the three variants of the general solution of the Liouville equation
and which are parameterized by $f(x),\cX(x)$ and~$g(y),\cY(y)$.
The three solutions of~\eqref{HillsTr} are
\begin{align*}
A_{\id} &=\Bigl[(x-y)\Bigl(\int_0^x\xi f(\xi)\Id\xi - \int_0^y\eta g(\eta)\Id\eta \Bigr)
   - \int_0^x\xi^2 f(\xi)\Id\xi - \int_0^y\eta^2 g(\eta)\Id\eta\\
{}&{}\qquad{}
   + xy\int_0^x f(\xi)\Id\xi + xy\int_0^y g(\eta)\Id\eta\Bigr]\bigr/(x+y),\\
A_{\sin} &=\int_0^x f(\xi)\sin2(\xi+y)\Id\xi
   +\int_0^y g(\eta)\sin2(x+\eta)\Id\eta\\
{}&{}\qquad{}
   -2\cot(x+y)\Bigl[\int_0^x f(\xi)\sin^2(\xi+y)\Id\xi
      +\int_0^y g(\eta)\sin^2(x+\eta)\Id\eta\Bigr],\\
A_{\sinh} &=\int_0^x f(\xi)\sinh(\xi+y)\cosh(\xi+y)\Id\xi
   +\int_0^y g(\eta)\sinh(x+\eta)\cosh(x+\eta)\Id\eta\\
{}&{}\quad{}
   -2\coth(x+y)\Bigl[\int_0^x f(\xi)\sinh^2(\xi+y)\Id\xi
      +\int_0^y g(\eta)\sinh^2(x+\eta)\Id\eta\Bigr];
\end{align*}
we recall that here $x=\cX$ and $y=\cY$ w.r.t.\ the original setting
of equations~\eqref{EqIra} and~\eqref{Hills}.

\begin{rem}\label{RemLiouDiag}
If the Liouville\/-\/type system~$\cE_\IL$
with $m$ unknown functions possesses a complete set of $m$~integrals, then it is solvable and its general solution depends on $2m$ arbitrary functions $f^i(x)$, $g^j(y)$, where $1\leq i,j\leq m$. The integrals determine the substitution from~$\cE_\IL$ to $m$ generators of $\ker\bar{D}_y$ and $m$ analogous elements of~$\ker\bar{D}_x$, which become the $m$-tuples $\{\phi^i(x)\}$, $\{\gamma^j(y)\}$ on any solution of~$\cE_\IL$. Note that $f^i(x)+g^i(y)$ and $\phi^i(x)+\gamma^i(y)$ determine $m$~solutions of
the wave equation~$\cE_\varnothing=\{s_{xy}=0\}$
(alternatively, one can get $2m$
solutions of the wave equations by potentiating the kernels).
Thus the hyperbolic Liouville\/-\/type systems can be defined as the diagrams
\begin{equation}\label{LDef}
\cE_{\varnothing}\xrightarrow{\text{sol}}\cE_\IL\xrightarrow{\text{int}}\cE_\varnothing,
\end{equation}
where the first arrow assigns general solutions of~$\cE_\IL$ to the
$m$~pairs of arbitrary functions and the second arrow is determined by
the integrals. In particular, diagram\eqref{LDef} leads to explicit
formulas of B\"acklund autotransformations for the nonlinear
systems~$\cE_\IL$.

Suppose further that the Euler\/--\/Lagrange Liouville\/-\/type system $\cEEL=\{\bE(\cL)=0\}$ is ambient w.r.t.\ the hierarchy of~$\cE'$. By construction, the Hamiltonian system~$\cE'$ specifies integrable Noether's symmetry flows on~$\cEEL$ and hence induces the dynamics $\boldsymbol{f}(x,t)$, $\boldsymbol{g}(y,t)$ that starts from the Cauchy data $\boldsymbol{f}(x,0)=\{f^i(x)\}$, $\boldsymbol{g}(y,0)=\{g^j(y)\}$. This way we obtain the evolutionary system~$\cE''$ upon $\boldsymbol{f}(x,t)$ or~$\boldsymbol{g}(y,t)$. The new system admits the Miura\/-\/type transformation to~$\cE'$. Indeed, it is given through the formulas for the general solution of~$\cEEL$. Hence the use of the auxiliary exactly solvable hyperbolic systems embeds~$\cE'$ in the triple $\cE''\to\cE'\to\cE$ of integrable equations; this diagram is composed by symmetries of diagram~\eqref{LDef} that defines the Liouville\/-\/type systems.
\end{rem}

\begin{example}\label{ExKN}
The potential modified KdV flow $V_t=-\tfrac{1}{2}V_{xxx}+V_x^3$ lifts onto the general solution $V=\tfrac{1}{2}\ln\bigl[\cX'(x)\cY'(y)/(\cX+\cY)^2\bigr]$ of the Liouville equation $V_{xy}=\exp(2V)$, resulting in $\cY_t=0$ and the Schwarz\/-\/KdV equation
\[
\cX_t=-\tfrac{1}{2}\cX_x\cdot\{\cX,x\}\equiv-\tfrac{1}{2}\cX_{xxx}+\tfrac{3}{4}{\cX_{xx}^2}\bigr/{\cX_x},
\]
which is Krichever\/--\/Novikov type.
\end{example}

Let us note that the classical Gardner deformations~$(\cE_\veps,\gm_\veps)$ can be found directly by using the Liouville\/-\/type systems 
but not passing to the systems in the image of the substitutions determined by the integrals. 
Namely, we suggest to find first the family of Liouville\/-\/type extensions $\cE_\IL^\veps$
that contains the ambient system~$\cEEL$ at $\veps=0$ and such that there is the contraction $\gm_\veps\colon\cE_\IL^\veps\to\cEEL$. This is done as follows. Clearly, the contraction extends the integrals $w$ for~$\cEEL$ onto the curve $\cE_\IL^\veps$ by the formula~$w^\veps=w[\gm_\veps]$. Then the conditions $\bar{D}_y(w[\gm_\veps])\doteq0$ satisfied on~$\cE_\IL^\veps$ correlate the known differential functions~$w$ with the admissible right\/-\/hand sides of~$\cE_\IL^\veps$ and the contractions $\gm_\veps$.
Then $\gm_\veps$~is inherited by the extensions $\cE_\veps$ of the integrable evolutionary flows~$\cE$ on~$\cEEL$. Thus we conclude that the integrals of the Liouville\/-\/type systems permit finding Miura's transformations \emph{to} systems of this type and symmetry flows on them, and not only \emph{from} them by inspecting the evolution of their integrals along symmetries.


\begin{example}\label{ExLioue}
Consider the extended potential modified KdV equation
(see~\eqref{mKdVe})
\[
\tV_t=-\tfrac{1}{2}\tV_{xxx}+\tV_x^3+{3\veps^2\tV_x\tV_{xx}}\bigr/{(1+4\veps^2\tV_x^2)}.
\]
The ambient hyperbolic equation
\begin{equation}\label{Lioue}
\cE_\IL^\veps=\bigl\{\tV_{xy}=\exp(2\tV)\cdot\sqrt{1+4\veps^2\tV_x^2}\bigr\}
\end{equation}
is not Euler\/--\/Lagrange if $\veps\neq0$ (one may expect this since the extended modified KdV equation~\eqref{mKdVe} looses the first Hamiltonian structure~$D_x$ if $\veps\neq0$).
The contraction $\gm^\veps\colon\cE_\IL^\veps\to\cEEL$ from~\eqref{Lioue} to the Liouville equation~$\cEEL=\{V_{xy}=\exp(2V)\}$ is given through
\begin{equation}\label{LioueLiou}
V=\tV-\tfrac{1}{2\veps}\arcsinh(2\veps\tV_x).
\end{equation}
Extension~\eqref{Lioue} and contraction~\eqref{LioueLiou} are obtained by the requirement that the equations~$\cE_\IL^\veps$ remain Liouville\/-\/type and their 
integrals are inherited by~$\gm_\veps$ from the integral $w=V_x^2-V_{xx}$ of the Liouville equation~$\cEEL$.

The scalar hyperbolic equation~\eqref{Lioue} appeared in entry~$4$ of the classification~\cite{SokolovUMN}. We also note that the curve~$\cE_\IL^\veps$, which contains a representative of the class~\cite[\S7]{SokolovUMN}
$\tV_{xy}=S(\tV)\cdot\sqrt{1-\tV_x^2}$ with $S''+\const^2 S=0$, is an extension of $\tV_{xy}=S(\tV)$ onto~$\veps\in\mathrm{i}\BBR$. We conclude that the Laplace invariants of $\cE_\IL^\veps$ satisfy the  deformation of the nonperiodic Toda chain and hence the deformed chain can not be periodic.
\end{example}

\section{The adjoint systems and extended Magri schemes}\label{SecAdj}
\noindent%
We have showed that not all Gardner's deformations are obtained using a symmetry of a system at hand, thus giving rise to the Gardner cohomology. The second \emph{ad hoc} principle which we claim is not true is that the curves~$\cE_\veps$ interpolates the initial systems~$\cE$ with prescribed modified equations. This assumption was postulated in~\cite{MathieuN=2} for an $N=2$ Super\/-\/KdV equation and implied a sequence of obstructions to the (still unknown) deformation. We note that the (Kaup-)\/Boussinesq systems in Examples~\ref{ExKBe} and~\ref{ExBouse} do possess the modified analogs (see~\cite{TMPhGallipoli,ConstCurv}), but this property has no relation to the respective Gardner deformations.
For example, the third order extension~\eqref{Bouse} can not be obtained from the Boussinesq system by any interpolation towards the second order modified equation, which was studied in~\cite{TMPhGallipoli}.

In this section we
describe a technique that extracts new completely integrable systems
from the Gardner deformations or, more generally, from
certain families~$\{\cE_\veps\}$ of completely integrable systems
that may not admit the contractions~$\gm_\veps\colon\cE_\veps\to\cE$.
From now on, we make a technical assumption that the extensions~$\cE_\veps$ are polynomial in the parameter~$\veps$.

Consider a curve $\{\cE_\veps\}$ of completely integrable systems (e.g., obtained via a Gardner's deformation); by assumption, for every~$\veps$ the system~$\cE_\veps$ admits the hierarchy~$\gA_\veps$ of higher flows that are polynomial in~$\veps$.
The \emph{adjoint system}~$\cE'$ and its higher flows are obtained by
isolating the coefficients of the highest powers of~$\veps$ in
$\cE_\veps$ and its symmetries, respectively.

Two types of the adjoint systems are further recognized.
Suppose $\gA$~is the hierarchy of higher symmetries for a system~$\cE$.
Let~$A_i$, $i\geq1$ be the Hamiltonian structures for~$\gA$ and
let~$A_{i,\veps}=A_i+\dots+\veps^{k_i}\cdot A_i'$
be a polynomial deformation of~$A_i$ such that
$A_{i,\veps}$~remains a Hamiltonian operator for all~$\veps$
(clearly, $A_i'$ is a Hamiltonian operator and is connected with
the original structure~$A_i$ by the homotopy conditions
$\lshad A_{i,\veps},A_{i,\veps}\rshad=0$, here~$\lshad\ ,\ \rshad$ is the Schouten bracket, see~\cite{Opava}).
Assume that the extensions~$A_{i,\veps}$ and~$A_{j,\veps}$ of the Hamiltonian operators are compatible for some $i$ and~$j$
(that is, $\lshad A_{i,\veps}, A_{j,\veps}\rshad=0$).
Finally, suppose that for each flow $\vph_{j,\veps}=A_{j,\veps}(\bE(\cH_{j,\veps}))$ there is the Hamiltonian~$H_{j+1,\veps}$ such that $\vph_{j,\veps}=A_{i,\veps}(\bE(\cH_{j+1,\veps}))$; this is realized
if the Poisson cohomology w.r.t.\ $A_{i,\veps}$ is trivial, see~\cite{Getzler,Lstar}.
The Magri scheme generated by the pair $A_{i,\veps}$, $A_{j,\veps}$
extends the bi\/-\/Hamiltonian hierarchy~$\gA$ to the family of hierarchies~$\gA_\veps$, which are polynomial in~$\veps$.
Consider the coefficients~$\vph'_k$ at the higher powers of~$\veps$ in the flows of~$\gA_\veps$. The flows~$\vph'_k$ are bi\/-\/Hamiltonian w.r.t.\ $A_i'$ and~$A_j'$ and constitute the \emph{adjoint hierarchy $\gA'$ of first type}.

\begin{example}
The extended KdV equation~\eqref{KdVe}
is bi\/-\/Hamiltonian w.r.t.\ the operator
$A_{1,\veps}=D_x$ and the weakly nonlocal structure
\[
{A}_{2,\veps}=-\tfrac{1}{2}D_x^3 + 2\tu\,D_x + \tu_x
+2\veps^2\cdot\bigl(\tu^2\,D_x + \tu\tu_x - \tu_x\,D_x^{-1}\circ \tu_x\bigr).
\]
This operator can be found by factorization of the recursion for~\eqref{KdVe} or by a direct calculation using the technique of~\cite{Lstar}. Isolating the flow at $\veps^2$ in~\eqref{KdVe}, we obtain the adjoint KdV equation $\dot{u}=u^2u_x$. Note that the scaling weights are not uniquely defined for the dispersionless modified KdV; they are fixed by switching on the dispersion in~\eqref{mKdV}.
\end{example}

Further, suppose $A_{i,\veps}=A_i+\veps^{k_i} A_i'$, then the Hamiltonian operators $A_i$ and $A_i'$ are compatible. Again, if every flow $\vph_\alpha=A_i'(\bE(\cH_\alpha))$ that belongs to the image of~$A_i'$ (or, vice versa, of~$A_i$) can be resolved w.r.t.\ the Hamiltonian~$\cH_{\alpha+1}$ such that $\vph_\alpha=A_i(\bE(\cH_{\alpha+1}))$ (respectively, $\vph_\alpha$ belongs the the image of~$A_i'$),   
then the extended Hamiltonian operator~$A_{i,\veps}$ generates the \emph{adjoint hierarchy of second type}.

\begin{example}
The adjoint Kaup\/--\/Boussinesq equation is (we omit the tilde signs)
\begin{equation}\label{AdjBous}
\left\{
\begin{aligned}
\dot{u}&= uu_{xx} + u_x^2 + u_xv + uv_x,\\
\dot{v}&= - \bigl(2u_xu_{xx} +uu_{xxx} + u_xv_x + uv_{xx} - vv_x\bigr).
\end{aligned}
\right.
\end{equation}
This system is bi\/-\/Hamiltonian w.r.t.\ the compatible
local operators $A_1=\left(\begin{smallmatrix}0 & D_x \\ D_x & 0
\end{smallmatrix}\right)$ and
\[
A_1'=\left(\begin{array}{cc}
0 & D_x\circ u_x + D_x\circ v\mathstrut \\
u_x D_x + v D_x &
  -u_{xx}D_x - D_x\circ u_{xx} - v_xD_x - D_x\circ v_x
\end{array}\right).
\]
Indeed, we have $\binom{\dot u}{\dot v} =
  A_1\bigl(\bE\bigl(\int\tfrac{1}{2}(uu_x^2+2uu_xv+uv^2)\,\Id
     x\bigr)\bigr) =
  A_1'\bigl(\bE\bigl(\int uv\,\Id x\bigr)\bigr)$.
Hence equation~\eqref{AdjBous} is a completely integrable
adjoint system of second type.

System~\eqref{AdjBous} admits the Galilean symmetry
$x\mapsto x+\lambda t$, $u\mapsto u$, $v\mapsto v+\lambda$; the original Kaup\/-\/Boussinesq equation is invariant under the transformation $x\mapsto x+\lambda t$, $u\mapsto u+\lambda$, $v\mapsto v$. Hence the shift is transferred from~$u$ to~$v$ by passing from~\eqref{KB} to its adjoint system~\eqref{AdjBous}.
The extensions~\eqref{KBe} of the Kaup\/-\/Boussinesq equation are not Galilei\/-\/invariant, and we also note that family~\eqref{KBe} is not obtained by the action of the Galilean symmetry.
\end{example}

\begin{rem}
The notions of Gardner's deformations~\eqref{GDiag} and extensions~$\gA_\veps$ of the Magri
schemes intersect if there is the contraction
$\hat{\gm}_\veps\colon\gA_\veps\to\text{sym}\,\cE$ from the symmetry hierarchy of the extension~$\cE_\veps$ and
if the first Hamiltonian structure~$A_{1,\veps}$ of the extended Magri scheme~$\gA_\veps$ is ~$A_{1,\veps}=\text{const}\cdot D_x$ as everywhere above; note that the results of~\cite{Getzler} on the triviality of the Poisson cohomology are valid in this case.
Generally, the extension $\gA_\veps$ of the Magri scheme~$\gA$ for an equation~$\cE$ may have no contraction~$\hat{\gm}_\veps$ to~$\cE$ and its symmetries. Hence the task of reconstructing the Hamiltonians of $\gA$ and the adjoint hierarchies~$\gA'$ becomes nontrivial.

The distinction between the Gardner deformations and
the extensions of the Magri schemes allows to advance with respect to the approach based on extending the Lax operators~\cite{KuperIrish} that mixes the two theories. Owing to this distinction in
Example~\ref{ExBouse} we improved a result of~\cite{KuperIrish}.
\end{rem}

\paragraph*{Final remarks.}
The use of ambient Liouville\/-\/type Euler\/--\/Lagrange equations
allows to reveal a not obvious property of the evolutionary systems
under study, namely, the presence and form of
their pre\/-\/Hamiltonian structures in the sense
of~\cite{SokolovUMN}. For example, the potential modified KdV equation
$V_t=-\tfrac{1}{2}V_{xxx}+V_x^3$ and its higher flows, which are
Noether's symmetries of the Liouville equation $V_{xy}=\exp(2V)$, see
Example~\ref{ExKN}, belong to the image of the operator $
u_x+\tfrac{1}{2}D_x$.
Using the classification theorem for Noether's symmetries of the Liouville\/-\/type systems~\cite{TMPhGallipoli}, we assign the matrix extension of this operator 
to system~\eqref{EqIra} and the potential twice\/-\/modified Kaup\/--\/Boussinesq equation~\eqref{pm2KB} on it.
The bracket in the inverse image of the matrix operator extends the
bracket for the second Hamiltonian structure of
KdV~\cite{TMPhGallipoli,SokolovUMN}.
Algebraic properties of the pre\/-\/Hamiltonian operators and their
relation to integrable systems will be analyzed 
in a subsequent publication~\cite{Moreva}.

We have already mentioned, when recalling the approach to constructing families of equations~$\cE_\veps$ by using a symmetry of $\cE$ that does not lift to $\cE_{\veps_0}$ at some $\veps_0$, that the symmetry algebras of the modified systems~$\cE'$ are Lie subalgebras in $\sym\cE$; in particular, we have $\sym\cE_{\veps_0}\subseteq\sym\cE$. An analogous relation is established for the fundamental Lie algebras of differential equations~\cite{FundLieAlg}: the fundamental algebras of $\cE_\veps$ form a one\/-\/parametric family of Lie subalgebras in the fundamental Lie algebra of~$\cE$. Actually, the use of modified systems with a parameter is the most helpful way of calculating these structures for PDE. We emphasize that Gardner's deformations generate the families~$\cE_\veps$ together with the contractions $\gm_\veps\colon\cE_\veps\to\cE$ in the most regular manner.

The proposed definition of Gardner's deformations is motivated by their principal task, specification of recurrence relations for the Hamiltonians of (super-)\/equations
that are not deformable through the classical scheme.
It is very likely that the difficulties which arise in the open problem~\cite{MathieuOpen} on constructing Gardner's deformations for the $N=2$, $a=-2,1,4$ Super\/-\/KdV equations can be removed by using this approach. First, the presence of obstructions to a deformation, which is observed in~\cite{MathieuN=2}, may mean the existence of a Miura's mapping~$\tau$ from the system under study. Secondly, since the first Hamiltonian structure for $N=2$ Super\/-\/KdV with $a=4$ is $A_1=D_x$, 
a generalization of the method of~\cite{TMPhGallipoli} for super\/-\/equations is applicable.
Next,               
no interpolation with a modified system must be \emph{a priori} required for the Gardner deformations.
Finally, it is normal that the extensions~$\cE_\veps$ and the adjoint systems loose local Poisson structures but remain integrable, inheriting the Hamiltonians through the contractions. Thus we obtain a class of completely integrable systems which are not manifestly bi\/-\/Hamiltonian. This situation may be realized for the $N=2$ Super\/-\/KdV${}_{a=1}$ equation~\cite{MathieuOpen,MathieuN=2}. Hence, regarding this super\/-\/system as adjoint, we face the task of reconstructing the adjoint to adjoint, that is, the non\/-\/extended super\/-\/equation~$\cE$. This will be the object of another paper.

\paragraph*{Acknowledgements.}
The author thanks V.\,V.\,Sokolov, V.\,N.\,Roubtsov,
I.\,S.\,Kra\-sil'\-sh\-chik, and A.\,Karasu for
helpful discussions and also is grateful to the referee for important comments and suggestions.
A part of this research was done while the author was visiting at Max Planck Institute for Mathematics (Bonn) and at~$\smash{\text{IH\'ES}}$.
This work is supported in part by~MPIM and~$\smash{\text{IH\'ES}}$.


\begin{thebibliography}{99}\normalsize

\bibitem{Gardner}
\by{Miura R.M., Gardner C.S., and Kruskal M.D.} (1968)
   Korteweg\/--\/de Vries equation and generalizations.~II.
   Existence of conservation laws and constants of motion,
\jour{J.~Math.\ Phys.} \vol{9}, 1204--1209.

\bibitem{MathieuOpen}
\by{Mathieu P.} (2001) Open problems for the super KdV equations. B\"acklund and Darboux transformations. The geometry of solitons. \book{CRM Proc.\ Lecture Notes} \vol{29}, 326-334.

\bibitem{MathieuN=2}
\by{Labelle P., Mathieu P.} (1991) A new $N=2$ supersymmetric Korteweg\/--\/de Vries equation, \jour{J.~Math.\ Phys.} \vol{32} n.4, 923-927.

\bibitem{PamKale}
\by{Karasu A., Kiselev A.V.} (2006)
   Gardner's deformations of the Boussinesq equations,
\jour{J.~Phys.\ A\textup{:} Math.\ Gen.} \vol{39} n.37,
11453-11460.\ \texttt{arXiv:nlin.SI/0603029}

\bibitem{KuperIrish}
\by{Kupershmidt B.A.} (1983)
   Deformations of integrable systems,
\jour{Proc.\ Roy.\ Irish Acad.} \vol{A83} n.1, 45--74.

\bibitem{ThreeApp}
\by{Kersten P., Krasil'shchik I., Verbovetsky A.} (2004) Nonlocal constructions in the geometry of PDE, in: Symmetry in nonlinear mathematical physics. \jour{Prepr.\ Inst.\ Mat.\ Nats.\ Acad.\ Nauk Ukra\"\i ni, Mat.\ Zastos.} \vol{50}, 412--423.

\bibitem{Opava}
\by{Krasil'shshik I., Verbovetsky A.} (1998)
\book{Homological methods in equations of mathematical physics}.
Open Education and Sciences, Opava.

\bibitem{TMPhGallipoli}
\by{Kiselev A.V.} (2005) Hamiltonian flows on Euler-type equations,
\jour{Theor.\ Math.\ Phys.} \vol{144} n.1, 952-960.\ \texttt{arXiv:nlin.SI/0409061}

\bibitem{SokolovUMN}
\by{Zhiber A.V., Sokolov V.V.} (2001)
  Exactly integrable hyperbolic equations of Liouvillean type,
\jour{Russ.\ Math.\ Surveys} \vol{56} n.1, 61--101.

\bibitem{DSViniti84}
\by{Drinfel'd V.G., Sokolov V.V.} (1985)
 Lie algebras and equations of Korteweg-de Vries type,
\jour{J.~Sov.\ Math.} \vol{30}, 1975-2035.

\bibitem{BorisovZykov}
\by{Borisov A.B., Zykov S.A.} (1998) The dressing chain of discrete symmetries and the proliferation of nonlinear equations, \jour{Theor.\ Math.\ Phys.} \vol{115} n.2, 530-541.

\bibitem{Dubrovin}
\by{Carlet G., Dubrovin B., Zhang Y.} (2004) The extended Toda hierarchy, \jour{Mosc.\ Math.~J.} \vol{4} n.2, 313-332, 534;
\by{Dubrovin B., Zhang Y.} (1998) Bi\/-\/Hamiltonian hierarchies in $2D$ topological field theory at one\/-\/loop approximation, \jour{Commun.\ Math.\ Phys.} \vol{198} n.2, 311-361.

\bibitem{Getzler}
\by{Getzler E.} (2002)
   A Darboux theorem for Hamiltonian operators
   in the formal calculus of variations,
\jour{Duke Math.~J.} \vol{111}, 535--560.

\bibitem{Lstar}
\by{Kersten P., Krasil'shchik I., Verbovetsky A.} (2004) Hamiltonian operators and $\ell^*$-coverings, \jour{J.~Geom.\ Phys.} \vol{50} n.1--4, 273--302.

\bibitem{MathieuN1}
\by{Mathieu P.} (1988)
   Supersymmetric extension of the Korteweg\/--\/de Vries equation,
\jour{J.~Math.\ Phys.} \vol{29} n.11, 2499--2506.

\bibitem{Kiev2005}
\by{Kiselev A.V., Wolf T.} (2006)
Supersymmetric representations and integrable fermionic extensions
of the Burgers and Boussinesq equations,
\jour{SIGMA~-- Symmetry, Integrability
and Geometry\textup{:} Methods and Applications} \vol{2}
n.30, 19~p.\ \texttt{arXiv:math-ph/0511071}

\bibitem{Gerstenhaber}
\by{Gerstenhaber M., Schack S.D.} (1988) Algebraic cohomology and deformation theory, in: Deformation theory of algebras and structures and applications (M.~Gerstenhaber and M.~Hazelwinkel, eds.) Kluwer, Dordrecht, 11--264.

\bibitem{DeformLiou}
\by{Kiselev A.V.} (2002)
   On B\"acklund autotransformation for the Liouville equation
\jour{Vestnik Moskovskogo Univ.\ Ser.~3 Fiz.\ Astr.} \vol{6} 22--26.


\bibitem{ConstCurv}
\by{Pavlov M. V.} (2002)
Integrable systems and metrics of constant curvature,
\jour{J.~Nonlin.\ Math.\ Phys.} \vol{9} Suppl.~1, 173--191.

\bibitem{Demskoi}
\by{Demskoi D.K.} (2007) 
On application of Liouville type systems to constructing
B\"acklund transformations,
\jour{J.~Nonlin.\ Math.\ Phys.}, 20~p.\ (to appear).

\bibitem{Moreva}
\by{Kiselev A. V., van de Leur J. W.} (2007)
Pre-Hamiltonian structures of integrable systems,
Proc.\ 2006 Twente Conf.\ on Lie groups,
The Netherlands (December 13--15, 2006), in preparation.

\bibitem{FundLieAlg}
\by{Igonin S.} (2006)
   Coverings and fundamental algebras for partial differential equations,
\jour{J.~Geom.\ Phys.} \vol{56} n.6, 939--998.

\end{thebibliography}
\end{document}